\documentclass[letterpaper, 10 pt, onecolumn]{IEEEtran} %
\IEEEoverridecommandlockouts %
\usepackage[english]{babel} %
\usepackage{graphicx} %
\usepackage{xcolor} %
\usepackage{caption}[2008/08/24] %
\usepackage{subcaption} %
\usepackage{xparse} %
\usepackage{xspace} %
\usepackage{grffile} %
\usepackage{csquotes} %
\usepackage{mathtools} %
\usepackage{amsthm} %
\usepackage{amssymb} %
\usepackage[locale=US, per-mode=fraction, binary-units=true]{siunitx} %
\usepackage{stmaryrd}
\usepackage{bbm}
\usepackage{bm} %

\usepackage{algorithm}
\usepackage{algorithmicx}
\usepackage[noend]{algpseudocode}

\usepackage{tikz} %
\usetikzlibrary{shapes,arrows,calc,intersections,backgrounds,positioning,plotmarks,decorations,decorations.text,automata}
\usetikzlibrary{babel,quotes,angles} %
\tikzset{>=latex} %

\usepackage{pgfplots} %
\pgfplotsset{compat=newest} %
\newlength\figureheight %
\newlength\figurewidth 	%

\usepackage{url} %
\usepackage[raiselinks=true, bookmarksnumbered=true, pdfusetitle]{hyperref} %
\hypersetup{colorlinks,	linkcolor=black, citecolor=black, urlcolor={blue!50!black}} %
\usepackage[capitalise]{cleveref} %
\crefname{equation}{}{} %

\newtheorem{example}{Example}
\newtheorem{assumption}{Assumption}

\newtheorem{definition}{Definition}
\newtheorem{claim}{Claim}

\renewcommand{\vec}[1]{\bm{#1}} %

\newcommand{\cvv}{\mathcal{V}}
\newcommand{\cee}{\mathcal{E}}
\newcommand{\cxx}{\mathcal{X}}
\newcommand{\cxxp}{\mathcal{X}^+}
\newcommand{\caa}{\mathcal{A}}
\newcommand{\cbb}{\mathcal{B}}
\newcommand{\cuu}{\mathcal{U}}

\newcommand{\RR}{\mathbb{R}}
\newcommand{\RRnn}{\RR_{\geq 0}}
\newcommand{\RRp}{\RR_{>0}}
\newcommand{\ZZ}{\mathbb{Z}}
\newcommand{\ZZp}{\ZZ_{> 0}}

\newcommand{\bool}{\mathbb{B}}
\newcommand{\true}{\top}
\newcommand{\false}{\bot}

\newcommand{\cs}{\vec{x}} %
\newcommand{\qss}{[\cxx]} %
\newcommand{\qssp}{[\cxx^+]} %
\newcommand{\qsspwi}{[\cxx^+_i]} %
\newcommand{\qs}{{[x]}} %
\newcommand{\qsp}{{[x^+]}} %
\newcommand{\qspwi}{{[x^+_i]}} %
\newcommand{\qbs}{{\vec{d}^x}} %
\newcommand{\qbsp}{\vec{d}^{x^+}} %
\newcommand{\qssi}[1]{[\cxx_#1]} %
\newcommand{\ci}{\vec{u}} %
\newcommand{\qis}{[\cuu]} %
\newcommand{\qi}{{[u]}} %
\newcommand{\qbi}{\vec{d}^u} %
\newcommand{\qisi}[1]{[\cuu_#1]} %

\newcommand{\tobits}{\mathsf{I}} %
\newcommand{\tovars}{\mathsf{V}} %
\newcommand{\overapprox}{{\mathsf{O}^{\system}}} %
\newcommand{\overapproxwi}{{\mathsf{O}^{\system}_i}} %
\newcommand{\trans}{\mathsf{T}} %
\newcommand{\transnext}{\trans'} %
\newcommand{\bfa}{\mathsf{BFA}} %
\newcommand{\ssa}{\mathsf{SSA}} %
\newcommand{\projwoi}{\mathsf{P}^{\system}} %
\newcommand{\proj}{\projwoi_i} %
\newcommand{\system}{\vec{f}} %

\newcommand{\edge}{edge~$e$\xspace}
\newcommand{\vertex}{vertex~$v$\xspace}
\newcommand{\timestep}{time step~$t$\xspace}
\newcommand{\ie}{i.e.\xspace}
\newcommand{\eg}{e.g.\xspace}
\newcommand{\comma}{, \;} %
\newcommand{\algin}{\qss \comma \qis \comma \qssp \comma \overapprox} %
\newcommand{\algout}{\trans} %
\newcommand{\boxpart}{[\caa]} %
\newcommand{\boxparti}{[\caa_i]} %
\newcommand{\boox}{[a]} %
\newcommand{\ith}{$i^{\text{th}}$\xspace} %
 
\begin{document}

\title{\LARGE \bf
Sparsity-Sensitive Finite Abstraction
} 

\author{Felix Gruber, Eric S. Kim, and Murat Arcak%
	\thanks{This work was supported in part by the German Academic Exchange Service, Heinrich and Lotte M\"uhlfenzl Foundation, NSF grant CNS-1446145, and the NSF Graduate Research Fellowship Program.}%
	\thanks{F. Gruber is with the Department of Electrical and Computer Engineering at the Technical University of Munich, 80333 M\"unchen, Germany.
		{\tt\small felix.gruber@tum.de}}%
	\thanks{E. Kim and M. Arcak are with the Department of Electrical Engineering and Computer Sciences at the University of California, Berkeley, CA 94720, USA.
		{\tt\small 	\{eskim, arcak\}@eecs.berkeley.edu}}%
}

\maketitle
\thispagestyle{empty}
\pagestyle{empty} %
\begin{abstract}

Abstraction of a continuous-space model into a finite state and input dynamical model is a key step in formal controller synthesis tools. 
To date, these software tools have been limited to systems of modest size (typically $\leq$ 6 dimensions) because the abstraction procedure suffers from an exponential runtime with respect to the sum of state and input dimensions.
We present a simple modification to the abstraction algorithm that dramatically reduces the computation time for systems exhibiting a sparse interconnection structure.
This modified procedure recovers the same abstraction as the one computed by a brute force algorithm that disregards the sparsity.
Examples highlight speed-ups from existing benchmarks in the literature, synthesis of a safety supervisory controller for a 12-dimensional and abstraction of a 51-dimensional vehicular traffic network. 

\end{abstract}
\section{Introduction}

A variety of software tools have been developed in the last decade to automatically synthesize controllers that enforce a formal specification. 
A common first step for symbolic controller synthesis requires abstracting the control system with a continuous state and input space into one with discrete states and inputs.
However, this abstraction step has been cited as a key bottleneck in benchmarks for higher-dimensional systems \cite{Nilsson2016}\cite{weber2016optimized}.
Many abstraction procedures implemented in current tools traverse the state space in a brute force manner and suffer from an exponential runtime with respect to the sum of state and input dimensions. 
The majority of continuous-space systems however contain a coordinate structure, where the governing equation for one state variable depends only on a small subset of other variables.

We present a simple modification to the brute force algorithm that takes advantage of sparsely interconnected structures. 
The modified abstraction algorithm judiciously looks at lower dimensional sub-spaces while ignoring those sub-spaces that do not affect the corresponding coordinate.
The proposed procedure returns the same abstraction of the control system as the brute force algorithm while exploiting the structure.
Our method requires only a linear number of operations with respect to the number of state dimensions, but this number grows exponentially with respect to a new sparsity parameter.
For control systems that exhibit a sparse interconnection structure, this introduced parameter is small compared to the sum of state and input dimensions.

We provide two examples to highlight the practical speed-ups that can be attained with this modified abstraction procedure. 
The first is a bicycle example that is a common benchmark among controller synthesis tools.
The second example is an arterial traffic corridor that is actuated with signalized intersections. These benchmarks confirm that the abstraction time and size of the abstraction encoding grow roughly linearly with the traffic network length.

\subsection{Literature Review}

Abstraction bottlenecks have motivated research into compositional synthesis \cite{Kim2015}\cite{Dallal2016} and construction of abstractions \cite{tabuada2001compositional}\cite{rungger2016compositional}. 
These techniques leverage a system decomposition and reason over interconnecting variables among sub-systems. 
Therefore, our abstraction procedure can complement compositional techniques by also taking advantage of sparsity \emph{within} a component.

Similar to our sparsity-sensitive finite abstraction algorithm, multiple procedures have appeared in the model checking and verification literature over the last decades. 
Finite abstractions of Markov processes have been constructed by leveraging independence assumptions on the underlying process noise \cite{Soudjani2015DBN}. 
In an application to symbolic model checking of sequential Boolean circuits, the abstraction encoding of a monolithic circuit has been decomposed implicitly in abstractions for each logical component independently \cite{Burch1991Partition}.
Whereas the literature above focuses on applications to model checking and verification, in this paper we construct finite abstractions with controller synthesis as end goal. 

\subsection{Software Tools} \label{subsec_software_tools}

In recent years, the breadth of abstraction and corresponding controller synthesis software tools has grown to include a greater variety of dynamical systems and classes of specifications. 
Examples of recent tools include CoSyMA \cite{mouelhi2013cosyma}, PESSOA \cite{Mazo2010}, and SCOTS \cite{Rungger2016}.

At their core, each of these tools relies on a system relation between the original and the abstract control system. 
Efficient algorithms and data structures from the formal methods community are then leveraged to enforce the specifications for the abstracted system.
The obtained controller of the finite input and state representation is then refined to the original system with continuous input and state spaces. 
The correctness of this controller design is guaranteed by relating the finite abstraction with the original control system via a system relation \cite{Tabuada2009}\cite{Reissig2016}.
In this paper, we utilize feedback refinement relations to construct controllers which require quantized state information only \cite{Reissig2016}.
However, the content in our paper can also accommodate other abstraction formalisms.
\section{Preliminaries}

Given two sets $\caa$ and $\cbb$, let $|\caa|$, $2^\caa$ and $\caa \times \cbb$ respectively represent $\caa$'s cardinality, power set (set of all subsets), and Cartesian product with $\cbb$. 

Let $\RR^n$ represent a $n$-dimensional Euclidean space. Given a vector element $\vec{x} \in \RR^n$, $x_i$ represents its \ith component for $i \in \{1,\ldots,n\}$. 
When convenient, we explicitly express $\vec{x}$'s coordinates as an ordered tuple $(x_1, \ldots, x_n)$.
Given a pair of vectors $\vec{x}, \vec{y} \in \RR^n$, let $\llbracket \vec{x}, \vec{y} \rrbracket $ denote the hyper-rectangle $[x_1,y_1] \times \ldots \times [x_n, y_n]$ and $\vec{z} + \llbracket \vec{x}, \vec{y} \rrbracket$ is the same except with an offset $\vec{z} \in \RR^n$. 
A set $\boox \subset \RR^n$ is a \emph{box} if there exists  $\vec{x}, \vec{y} \in \RR^n$ such that $\boox = \llbracket \vec{x}, \vec{y} \rrbracket$.
The vectors $\vec{x}$ and $\vec{y}$ are the lower and upper bound of $\boox$, respectively.

A set $\boxpart = \prod_{i=1}^n \boxparti$ is a \emph{box partition} of $\caa \subset \RR^n$, if each $\boxparti$ is an interval partition, \ie, $\boxpart$ inherits a coordinate structure and the union of all $\boox \in \boxpart$ forms a cover of $\caa$.
A non-empty intersection between two elements in $\boxpart$ may only occur if they are adjacent boxes, and such an intersection has zero measure. 
Therefore, a box partition is a generalization of uniform space partitions, where the boxes do not have to be all congruent to each other.

Every box $\boox \in \boxpart$ can be uniquely identified with a vector of indices $\vec{d}^a \in \ZZp^n$, where $d^a_i$ is the corresponding index of the \ith coordinate dimension.
Let $\tobits \colon \boxpart \to \ZZp^n$ be an invertible map from an element of a box partition to its corresponding vector of indices. 

The set of Booleans $\bool = \{\true, \false\}$ consists of two elements, Boolean true $\true$ and false $\false$. 
Let $\neg$, $\land$ and $\lor$ denote the logical negation, conjunction and disjunction, respectively. 
\section{Construction of Abstractions}

In this section, we present the brute force algorithm which constructs an abstraction by traversing the whole state and input space. 
Subsequently, an overview of existing methods to compute a reachable set over-approximation from a continuous-space control system is given.

\subsection{Control System}

A discrete-time, continuous-space \emph{control system} is given by a system of update equations
\begin{equation}
	\label{eqn_dynamics}
	\cs^+ = \system (\cs,\ci) = 	
	\begin{pmatrix}
		f_1 (\cs, \ci) \\
		\vdots \\
		f_n (\cs, \ci)
	\end{pmatrix}, 
\end{equation}
where $\cs^+\in \RR^n$ represents the state at the next time step. 
The function $\system \colon \cxx \times \cuu \to \cxxp$ is defined over a state space $\cxx \subset \RR^n$ and input space $\cuu \subset \RR^m$ and maps to the space of next states $\cxxp \equiv \cxx$, which are subsets of a Euclidean space of appropriate dimensions.

\begin{assumption}
	The state space $\cxx$ and input space $\cuu$ are bounded hyper-rectangles. 
\end{assumption}

By construction, $\qss$, $\qis$, $\qssp$ are finite and all corresponding box elements $\qs$, $\qi$, $\qsp$ can be represented as index vectors.
Moreover, all three box partitions inherit a coordinate structure, for instance $\qssp = \prod_{i=1}^n \qsspwi$.

\subsection{Finite Abstractions}

\begin{definition}
	A \emph{finite abstraction} of a control system is a quadruple $(\qss, \qis, \qssp, \trans)$, where $\qss$, $ \qis$, $\qssp$ are associated box partitions of $\cxx$, $\cuu$, $\cxxp$, and $\trans \colon \tobits(\qss) \times \tobits(\qis) \times \tobits(\qssp) \to \bool$ is a Boolean function that encodes valid transitions. 
\end{definition}

A transition from $\qbs \in \tobits(\qss)$ to $\qbsp \in \tobits(\qssp)$ under input $\qbi \in \tobits(\qis)$ is valid if and only if $\trans (\qbs,\qbi,\qbsp) = \true$. 
Thus, the pre-image $\trans^{-1}(\top) \subseteq \tobits(\qss) \times \tobits(\qis) \times \tobits(\qssp)$ is a set that encodes all valid transitions. 
Although the information stored in the transition function $\trans$ and the set of valid transitions $\trans^{-1}(\top)$ are equivalent, we opt to use the symbolic function representation because it can easily be compressed, as shown by the following example.

\begin{example}[Valid Transition Representations]
	Consider the transition function $\trans(\qbs,\qbi,\qbsp) = \top$ if and only if $\qbs = \qbsp$, which encodes stationary dynamics and no effect from control inputs.
	If $k$ and $l$ are the total number of bits used to represent an element of $\qss$ and $\qis$, respectively, then enumerating all elements of $\trans^{-1}(\top)$ in a sparse matrix lookup table requires $2^{k+l}$ non-zero entries. 
	On the other hand, $\trans$ can also be represented as Boolean function with $k$ clauses
	\begin{equation} 
		\trans (\qbs,\qbi, \qbsp) = \bigwedge_{i=1}^k \left(d^x_i == d^{x^+}_i \right),
	\end{equation}
	where the \ith clause encodes equivalence between the \ith index of $\qbs$ and $\qbsp$ for $i \in \{ 1, \ldots, k \}$.
\end{example}

\cref{sec_examples} presents further examples where Boolean function representations allow us to construct abstractions with more than $\num{E+60}$ valid transitions with as little as hundreds of kilobytes.

\subsection{Brute Force Abstraction}

A core subroutine in each of the controller synthesis tools presented in \cref{subsec_software_tools} is the computation of a finite abstraction from the continuous-space dynamics. 
Discrete states and inputs correspond with non-empty boxes in the continuous space. 
Transitions among discrete states $\qs$ under inputs $\qi$ are computed by first executing \cref{eqn_dynamics} from a representative state $\cs \in \qs$ and input $\ci \in \qi$, then adding additional transitions to account for the quantization error.
Since the computation of exact reachable sets of boxes $\qs$ under input $\qi$ is too difficult, the finite abstraction over-approximates the behavior of the original continuous-space control system. 

Let the reachable set over-approximation function of the control system be given by $\overapprox \colon \qss \times \qis \to 2^{\qssp}$ .
It maps a state and input box to a set of next state boxes.
We will provide further details about efficient over-approximation techniques for different classes of dynamical systems in \cref{subsec_reachset_overapprox}.

We assume that an over-approximation function $\overapprox$ is given to the brute force abstraction algorithm, which computes the transition function $\trans$ iteratively.
Since $\trans$ is a function with inputs, we introduce as placeholders some free variables $\tovars(\qss)$, $\tovars(\qis)$ and $ \tovars(\qssp)$. 
For instance, let $\tovars(\qss)$ be a map from a state box partition~$\qss$ to a vector of corresponding free variables.
The brute force abstraction procedure is given by \cref{alg_brute_abs}, where the double equal sign denotes component-wise equality. 
\begin{algorithm}
    \caption{Brute Force Abstraction $(\bfa)$}
	\label{alg_brute_abs}
    \begin{algorithmic}[1] %
	    \Function{$\bfa \left( \algin \right)$}{}
			\State $\trans \gets \false$
		    \ForAll{$\qs \in \qss$} 
		        \ForAll{$\qi \in \qis$} 
			    \State $\transnext \gets \false$
			    	\ForAll{$\qsp \in \overapprox( \qs, \qi )$}  
						\State $\transnext \gets \transnext \lor (\tovars(\qssp) == \tobits(\qsp))$
  			        \EndFor
  			        \State $\transnext \gets \transnext \land (\tovars(\qss) == \tobits(\qs))$
   			        \State $\transnext \gets \transnext \land (\tovars(\qis) == \tobits(\qi))$
   			        \State $\trans \gets \trans \lor \transnext$
		        \EndFor
	        \EndFor
	        \State \Return $\algout$
        \EndFunction
    \end{algorithmic}
\end{algorithm}
This algorithm iterates over all elements of $\qss$ and $\qis$, which can be seen as a nested for loop of depth equal to the sum of the state and input dimensions.
Thus, the number of discrete states and inputs grows exponentially with each additional dimension, \ie, obtaining a finite abstraction becomes an intractable problem for higher-dimensional systems.

\subsection{Reachable Set Over-approximations} \label{subsec_reachset_overapprox}

In \cref{alg_brute_abs}, we assumed that a method $\overapprox$ to over-approximate reachable sets is given.
There exist multiple techniques to over-approximate the system dynamics, tailored for different classes of systems.
In the following, we briefly present a few of them.

Let the two boxes $\qs = \llbracket \vec{a}^x, \vec{b}^x \rrbracket$ and $\qi = \llbracket \vec{a}^u, \vec{b}^u \rrbracket$ be the inputs of $\overapprox$.
An efficient estimate of the reachable set of nonlinear, monotone dynamical systems is presented in \cite{moor2002abstraction}.
The authors consider discrete-time systems that are monotone with respect to the non-negative convex cone $\RRnn^n$.
The over-approximation function for this class is
\begin{equation}
	\label{eqn_monotone_over}
	\overapprox = \left\lbrace \qsp \in \qssp \Big| \qsp \cap \llbracket \system (\vec{a}^x, \vec{a}^u), \system (\vec{b}^x, \vec{b}^u) \rrbracket \neq \emptyset \right\rbrace .
\end{equation}
Monotone systems are a subclass of mixed monotone systems, for which there exists a generalized procedure that accounts for increasing or decreasing components of $\system$ \cite{coogan2015efficient}.

An over-approximation method for systems with perturbed measurements is presented in \cite{Reissig2016}. 
A local error bound $\vec{\beta}$ is introduced, which bounds the uncertainty in each state coordinate independently based on the current $\qs$ and $\qi$.
Let the centers of the state and input box be $\vec{c}^x = \frac{1}{2}(\vec{b}^x - \vec{a}^x)$ and $\vec{c}^u = \frac{1}{2}(\vec{b}^u - \vec{a}^u)$, respectively. 
Then
\begin{equation}
	\label{eqn_bound_over}
	\overapprox = \left\lbrace \qsp \in \qssp \Big| \qsp \cap \left( \system (\vec{c}^x, \vec{c}^u) + \llbracket -\vec{\beta}, \vec{\beta} \rrbracket \right) \neq \emptyset \right\rbrace
\end{equation}
is the corresponding over-approximation function for this class of dynamical systems.
\section{Sparsity-Sensitive Abstraction for Discrete-Time Systems} \label{sec_DT_Locality}

In this section, we propose modifications of \cref{alg_brute_abs} for discrete-time, sparsely interconnected control systems.
The brute force algorithm is suited for systems where a change in each coordinate affects the next states in all other coordinates. 
However, when a coordinate update is independent of many other coordinates, then redundant computations are contained in \cref{alg_brute_abs}'s nested for loops. 
Thus, we can compute finite representations of systems more efficiently by exploiting this sparse interconnection structure.

\begin{example}[Discrete-Time Redundant Iterations] \label{example_dis_redundant}
	Consider the following system consisting of three state and two input variables
	\begin{equation}
		\begin{pmatrix}
			x_1^+ \\
			x_2^+ \\
			x_3^+
		\end{pmatrix} = 
		\begin{pmatrix}
			f_1(x_1, x_3, u_1, u_2) \\
			f_2(x_2, x_3, u_1, u_2) \\
			f_3(x_3, u_1, u_2)
		\end{pmatrix} ,
	\end{equation}
	where each coordinate update depends on at most four instead of all five possible variables. 
	Then \cref{alg_brute_abs}'s for loop iteration over independent variables provides no further information on the next state's component.
	Thus, instead of dealing with three 5-dimensional update equations, we can consider only two 4-dimensional and one 3-dimensional problems.
\end{example}

\subsection{Dependency Graph}
\label{subsec_dependency_graph}

Before we propose modifications to \cref{alg_brute_abs}, we illustrate system dependencies as a directed graph, called \emph{dependency graph} subsequently. 

\begin{figure}[htbp]
	\centering
	%\tikzsetnextfilename{BicycleGraph}  
	%
\begin{tikzpicture}[->, very thick]

\node[regular polygon, regular polygon sides=5, minimum size=3cm] at (5*4,0) (center) {};
\node[state] (x1) at (center.corner 2) {$x_1$};
\node[state] (x2) at (center.corner 1) {$x_2$};
\node[state] (x3) at (center.corner 5) {$x_3$};
\node[state] (u1) at (center.corner 3) {$u_1$};
\node[state] (u2) at (center.corner 4) {$u_2$};

\draw (u1) -- (x1);
\draw (u1) -- (x2);
\draw (u1) -- (x3);
\draw (u2) -- (x1);
\draw (u2) -- (x2);
\draw (u2) -- (x3);
\draw (x3) -- (x1);
\draw (x3) -- (x2);

\path[loop above]
	(x1) edge ()
	(x2) edge ()
	(x3) edge ()
;

\end{tikzpicture}
 	\caption{Dependency graph for control system in \cref{example_dis_redundant}.}
	\label{fig:BicycleGraph}
\end{figure}
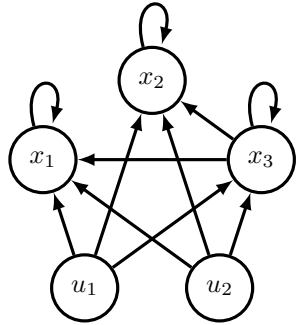

Each vertex of the dependency graph $v_i$ for $i \in \{ 1, \ldots, n+m \}$ represents a state or input variable. 
If the state update equation associated with vertex~$v_i$ depends on $v_j$, a directed edge from $v_j$ to $v_i$ is added. 
Consequently, the indegree of a vertex is equal to the number of directly dependent states and inputs.
The dependency graph for the control system in \cref{example_dis_redundant} is illustrated in \cref{fig:BicycleGraph}.
If the indegree of a next state vertex is smaller than the sum of state and input variables, \ie if the next state variable does not depend on all possible variables, it is advantageous to exploit the sparse interconnection structure.

\subsection{Coordinate-Wise Decomposition}
\label{subsec_decomposition}

In order to exploit sparsely interconnected systems, we first define a vector of projections $\projwoi$ associated with $\system$. 
Let the \ith component $\proj$ of $\projwoi$ be a projection of a discrete box or corresponding box partition onto the lower dimensional domain of the $f_i$ component for $i \in \{ 1, \ldots, n \}$.
For instance, the associated box partition projections of the first component of $\system$ in \cref{example_dis_redundant} are given by $\projwoi_1 (\qss) = \qssi{1} \times \qssi{3}$ and $\projwoi_1 (\qis) = \qisi{1} \times \qisi{2}$, respectively.

Before we present our algorithm modifications, we need to make the following assumption, which is fulfilled for all reachable set over-approximation functions in \cref{subsec_reachset_overapprox}.
\begin{assumption} \label{ass_over_approx}
	The over-approximation function $\overapprox$ can be decomposed coordinate-wise. 
	This means that $\overapprox( \qs, \qi ) = \prod_{i=1}^n \overapproxwi \left( \proj(\qs), \proj(\qi) \right)$, where the over-approximation method associated with the \ith coordinate dimension is given by $\overapproxwi \colon \proj(\qss) \times \proj(\qis) \to 2^{\qsspwi}$ for $i \in \{ 1, \ldots, n \}$.
\end{assumption}

\subsection{Sparsity-Sensitive Abstraction}

We propose \cref{alg_DT_abs} below, which takes advantage of dependencies among state, input and next state variables by executing a lower dimensional brute force abstraction for each coordinate independently.
\begin{algorithm}
	\caption{Sparsity-Sensitive Abstraction $(\ssa)$}
	\label{alg_DT_abs}
	\begin{algorithmic}[1] %
		\Function{$\ssa \left( \algin \comma \projwoi \right)$}{}
			\State $\trans \gets  \true$
			\ForAll {$i \in \{1,\ldots, n\}$}
				\State $\transnext \gets \bfa \left( \proj(\qss) \comma \proj(\qis) \comma \qsspwi \comma \overapproxwi \right) $
				\State $\trans \gets \trans \land \transnext$
			\EndFor
			\State \Return $\algout$
		\EndFunction
	\end{algorithmic}
\end{algorithm}

\begin{claim}(Equivalence between Abstraction Algorithms)\label{claim_equiv}
	\cref{alg_brute_abs} and \cref{alg_DT_abs} return the same abstraction. 
\end{claim}

A proof is provided in the appendix.

Let $g = \max \big(|\qssi{1}|,\ldots, |\qssi{n}|,|\qisi{1}|,\ldots, |\qisi{m}| \big)$ be the maximum number of intervals along all dimensions.
Furthermore, let the new sparsity parameter $h \leq n + m$ be the maximum indegree of all vertices in the dependency graph. 
The sparsity parameter $h$ can also be computed based on the adjacency matrix~$A \in \RR^{(n+m) \times (n+m)}$ of the dependency graph.
An element $A_{ij}$ of the adjacency matrix is one if there exists an edge from vertex~$v_i$ to $v_j$ and zero otherwise.
The indegree of vertex~$v_i$ is simply given by the sum of all entries in column~$i$.
For instance, the maximum indegree of the graph in \cref{fig:BicycleGraph} is $4$.
Then, the brute force \cref{alg_brute_abs}'s $O(g^{n+m})$ complexity is reduced to $O(ng^h)$.
While \cref{alg_DT_abs} is still exponential, the value of the exponent is smaller than the sum of states and inputs for control systems with sparse dependency structures, \ie $h < n+m$.
\section{Sparsity-Sensitive Abstraction for Continuous-Time Systems} \label{sec_CT_Locality}

For continuous-time systems, the dynamics are given by a system of differential equations. 
An abstraction, \ie a finite input and state representation, is associated with the corresponding sampled system with time quantization parameter $\tau \in \RRp$.
The system inputs are assumed to be constant for each sampling interval of length~$\tau$.
In \cref{sec_DT_Locality}, we have proposed \cref{alg_DT_abs} as a modification of \cref{alg_brute_abs}, which is valid for discrete-time systems. 
In discrete-time, a state at the next time step depends solely on the variables appearing in its update equation. 
This idea can be generalized to continuous-time systems, where a change in one variable can have an indirect effect on the time evolution of another one in infinitesimal amount of time. 
When the dynamics are given in continuous time, the notion of dependencies requires a slight modification, which we present in the following.

The dependency graph for the continuous-time case, similar to the discrete one in \cref{subsec_dependency_graph}, consists of vertices which represent a state or input variable. 
If the differential equation associated with vertex~$v_i$ depends on $v_j$, a directed edge from $v_j$ to $v_i$ is added, similar to the update equation in discrete-time. 
In contrast to the state update equations, the time evolution of a state in continuous-time always depends on the initial value.
Therefore, loops for all vertices which correspond to a state need to be added additionally in the dependency graph. 
\begin{example}[Continuous-Time Dependency Graph]\label{example_cont_redundant}
	Consider the following system of ordinary differential equations

	\begin{equation}
		\begin{pmatrix}
			\dot{x}_1 \\
			\dot{x}_2 \\
			\dot{x}_3
		\end{pmatrix} = 
		\begin{pmatrix}
			f_1(x_3, u_1) \\
			f_2(x_3, u_2) \\
			f_3(u_1, u_2)
		\end{pmatrix} .
	\end{equation}
	
	The associated dependency graph can be obtained from the graph in \cref{fig:BicycleGraph} by deleting the directed edges $(u_1, x_2)$ and $(u_2, x_1)$. 
\end{example}

The number of directly dependent variables of a vertex of the dependency graph is equal to the indegree.
Furthermore, the maximum indegree over all vertices corresponds to the sparsity parameter~$h$ in the discrete-time case.
Since there exist solely direct dependencies, we only need to consider walks of length one.
However, for continuous-time systems there can be indirect influences that are not immediately reflected in the system equations. 
For instance, consider a double integrator system where the position state depends indirectly on the incoming acceleration state via the velocity state.
In general, state vertex~$v_i$ depends on any vertex~$v_j$ if there exists a walk from $v_j$ to $v_i$.
The number of dependencies for each vertex can be computed via a breadth-first search for all state vertices.
Alternatively, we can compute these number for all vertices based on the adjacency matrix~$A$.
The element $\left(A^l \right)_{ij}$ corresponds to the number of walks of length~$l$ from vertex~$v_i$ to $v_j$.
Therefore, the number of vertices from which $v_i$ can be reached is equal to the number of non-zero elements of the \ith column of $\tilde{A} = \sum_{j = 0}^{n} A^j$.
We need to sum up to $n$, since the longest path from any vertex to another one is of length~$n$. 
A path is defined as a walk which does not traverse any vertex more than once.
Then, the sparsity parameter~$h$ for continuous-time systems is given by the maximum number of non-zero elements of all columns of $\tilde{A}$.

\section{Examples} \label{sec_examples}

We illustrate our sparsity-sensitive algorithm on two examples taken from the literature.
All computations are run on a single thread of an Intel\textsuperscript{\textregistered} Core\texttrademark\ i$7$-$5500$U ($\SI{3.00}{\GHz}$) with $\SI{8}{\giga\byte}$ RAM.
We use a modified version of the open-source tool SCOTS \cite{Rungger2016}, which represents transition functions, specifications and controllers as single binary decision diagrams (BDD) \cite{Bryant1986BDD}.

\subsection{Bicycle Model}

As a first example, we want to obtain an abstraction for the continuous-time, continuous-space bicycle model with corresponding parameters presented in \cite{Reissig2016}.
The position of the modeled vehicle in the $2$-dimensional plane and the orientation is given by the states $(x_1, x_2)$ and $x_3$, respectively.
The two controllable inputs $u_1$ and $u_2$ are the velocity and steering angle.
The vehicle dynamics are given by the following system of ordinary differential equations (ODEs)
\begin{equation} 
	\label{eqn_bicycle}
	\begin{pmatrix}
		\dot{x}_1 \\
		\dot{x}_2 \\
		\dot{x}_3
	\end{pmatrix}
	= 
	\begin{pmatrix}
		u_1 \cos (\alpha + x_3) / \cos(\alpha) \\
		u_1 \sin (\alpha + x_3) / \cos(\alpha) \\
		u_1 \tan (u_2)
	\end{pmatrix}
\end{equation}
with $\alpha = \arctan (\tan (u_2) / 2)$.

We can solve \cref{eqn_bicycle} analytically to obtain a discrete-time control system with sampling time~$\tau$, because the two inputs $u_1$ and $u_2$ are assumed to be constant for each sampling interval.
The corresponding dependency graph of the exact, discrete-time representation of the sampled vehicle system is illustrated in \cref{fig:BicycleGraph}.
Since the software tools presented in \cref{subsec_software_tools} solve \cref{eqn_bicycle} numerically to compute an abstraction, we can also use the same update rules of the employed ODE solver to get a discrete-time system.
Consequently, we obtain the identical transition function for both algorithms and can determine the computational speed-up.

The construction of the abstraction takes $\SI{2.3}{\second}$ based on our proposed \cref{alg_DT_abs}.
In contrast, it requires $\SI{104.0}{\second}$ to compute the identical transition function based on the brute force procedure implemented in \cite{Rungger2016}, while using the same over-approximation function \cref{eqn_bound_over}.
Therefore, we are able to speed-up the computation time by a factor of $45$, while the memory requirements are identical.
Due to its partial implementation in MATLAB\textsuperscript{\textregistered} for the nonlinear case, PESSOA \cite{Mazo2010} needs more than $\SI{e5}{\second}$ on a $\SI{2.40}{\GHz}$ machine to compute the abstraction \cite{Reissig2016}.

\subsection{Traffic Model}

The traffic model used in this example is taken from the modified, discrete-time cell transmission model presented in \cite{Kim2015}.
The traffic network can be represented as directed graph with a set $\cee$ of edges or links and a set $\cvv$ of vertices or junctions.
Every \edge is a directed arc from tail vertex $\tau (e) \in \{ \cvv \cup \emptyset \}$ to head vertex  $\eta (e) \in \cvv$.
By convention, an edge with $\tau (e) = \emptyset $ serves as entry point to the network and edges, which outflow the network, are not explicitly modeled.
Let
\begin{align}
	\cee^{\text{in}}_v &= \{e \in \cee \mid \eta (e) = v \} \\
	\cee^{\text{out}}_v &= \{e \in \cee \mid \tau (e) = v \}
\end{align}
denote the incoming, respectively, outgoing edges of \vertex.
The set of local edges to \edge is given by
\begin{equation}
	\cee^{\text{local}} (e) = \cee^{\text{in}}_{\tau (e)} \cup \cee^{\text{out}}_{\tau (e)} \cup \cee^{\text{out}}_{\eta (e)}.
\end{equation}

Each \edge$\in \cee$ has an associated link occupancy $x_e \in [0, \bar{x}_e]$ at time step $t \in \ZZ_{\geq 0}$, where the constant $\bar{x}_e$ denotes the maximum vehicular capacity of \edge.
Then, the continuous, $n$-dimensional state space of the traffic model is
\begin{equation}
	\label{eq:DimStateSpace}
	\cxx = \prod_{e \in \cee} [0, \bar{x}_e] \subset \RR^n, 
\end{equation}
where $n = |\cee|$. An \edge is actuated if the traffic signal at vertex~$\eta (e)$ permits outward flow from $e$ to $\cee^{\text{out}}_{\eta (e)}$.
The controllable input $u_e$ associated with \edge is given by
\begin{equation}
	u_e = 
	\begin{cases}
		1 & \text{if } \eta (e) \text{ actuates } e \text{ at \timestep} \\
		0 & \text{otherwise}
	\end{cases}
\end{equation}

The constant split ratio $\beta_{ek} \in [0, 1]$ denotes the fraction of vehicles flowing from \edge to edge $k \in \cee^{\text{out}}_{\eta (e)}$.
However, \edge can only send vehicles to $k$ if $k$ offers enough capacity.
For this purpose, the constant supply ratio $\alpha_{ek} \in [0, 1]$ denotes the capacity fraction of $k$ dedicated to $e$.
Since the supply is split among actuated incoming edges, it follows that for all $k \in \cee$ and time steps~$t$
\begin{equation}
	\sum_{e \in \cee^{\text{in}}_{\tau (k)}} \alpha_{ek} u_e = 1.
\end{equation}

In the following, we introduce the system dynamics based on the modified cell transmission model, which exhibits FIFO property.
As we will see, the outflow and state update function of \edge depends only on the corresponding states of edges in $\cee^{\text{out}}_{\eta (e)}$ and $\cee^{\text{local}} (e)$, respectively.
Thus, the traffic network exhibits a locality property, which we can exploit to compute abstractions.

The flow out of edge~$e \in \cee$ is given by
\begin{equation}
	\label{eq:outflow}
	f^{\text{out}}_e  = u_e \min \left( x_e \comma c_e \comma \min_{k \in \cee^{\text{out}}_{\eta (e)}} \left\{ \frac{\alpha_ {ek}}{\beta_{ek}} (\bar{x}_k - x_k) \right\} \right).
\end{equation}
The minimization in \cref{eq:outflow} implies that the flow of exiting vehicles of \edge cannot exceed the current link occupancy, the constant saturation flow~$c_e$ and the resulting supply offered by edges in $\cee^{\text{out}}_{\eta (e)}$. 
The saturation flow~$c_e$ is a model parameter, which is determined by, \eg, number of lanes and speed limits.
The inflow of \edge is denoted by
\begin{equation}
	f^{\text{in}}_e  = \sum_{k \in \cee^{\text{in}}_{\tau (e)}} \beta_{ke} f^{\text{out}}_k.
\end{equation}

Based on the principle of mass conservation, the discrete-time update equation of state $x_e$ associated with \edge is 
\begin{equation}
	\label{eq:stateupdate}
	x_e^+ = \min \left\{ \bar{x}_e \comma x_e - f^{\text{out}}_e + f^{\text{in}}_e + d_e\right\} ,
\end{equation}
where $d_e \in [0, \bar{d}_e ]$ represents a time-dependent, exogenous flow entering \edge and the constant parameter $\bar{d}_e$ is an upper bound.
It is clear based on \crefrange{eq:outflow}{eq:stateupdate}, that the state update equation of \edge depends only on the current states of edges in $\cee^{\text{local}} (e)$.
Therefore, we can reduce the computation time of the abstraction drastically by exploiting this sparse interconnection structure of the traffic network.

\subsection*{Arterial Traffic Corridor}

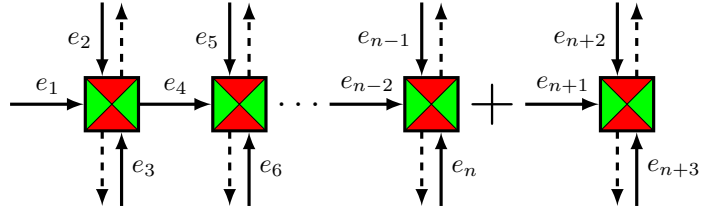
\begin{figure}[htbp]
	\centering
	%\tikzsetnextfilename{Network}  
	%
%
\begin{tikzpicture}[very thick]

\tikzset{vertex/.pic = 
	{
		\node[regular polygon, regular polygon sides=4, inner sep=0pt, outer sep=-0.35pt, minimum width=1.0cm, draw=black, fill=green] at (0,0) (-node){};
		\draw[fill=red, thin] (-node.corner 1) -- (-node.corner 2) -- (-node.center) -- cycle;						
		\draw[fill=red, thin] (-node.corner 3) -- (-node.corner 4) -- (-node.center) -- cycle;
	}
}

\def\topleft{110}
\def\topright{70}
\def\bottomleft{-110}
\def\bottomright{-70}

\pic [local bounding box=v_2] {vertex};
\pic [local bounding box=v_5, right = of v_2] {vertex};
\node [right = 0cm of v_5] (dots) {\Large $\ldots$};
\pic [local bounding box=v_n, right = of dots] {vertex};
\node [right = 0cm of v_n] (plus) {\Huge $+$};
\pic [local bounding box=v_n_3, right = of plus] {vertex};

\begin{scope}[->, dashed]
	\draw (v_2.\bottomleft) -- ++(0,-1);
	\draw (v_5.\bottomleft) -- ++(0,-1);
	\draw (v_n.\bottomleft) -- ++(0,-1);
	\draw (v_n_3.\bottomleft) -- ++(0,-1);
	\draw (v_2.\topright) -- ++(0,1);	
	\draw (v_5.\topright) -- ++(0,1);
	\draw (v_n.\topright) -- ++(0,1);
	\draw (v_n_3.\topright) -- ++(0,1);
\end{scope}

\begin{scope}[<-, shorten <=0.55pt]
	\draw (v_2.west) -- ++(-1,0) node[midway, above] {$e_1$};
	\draw (v_5) -- (v_2) node[midway, above] {$e_4$};
	\draw (v_n.west) -- ++(-1,0) node[midway, above] {$e_{n-2}$};
	\draw (v_n_3.west) -- ++(-1,0) node[midway, above] {$e_{n+1}$};
	\draw (v_2.\topleft) -- ++(0,1) node[midway, left] {$e_2$};	
	\draw (v_5.\topleft) -- ++(0,1) node[midway, left] {$e_5$};
	\draw (v_n.\topleft) -- ++(0,1) node[midway, left] {$e_{n-1}$};
	\draw (v_n_3.\topleft) -- ++(0,1) node[midway, left] {$e_{n+2}$};
	\draw (v_2.\bottomright) -- ++(0,-1) node[midway, right] {$e_3$};
	\draw (v_5.\bottomright) -- ++(0,-1) node[midway, right] {$e_6$};
	\draw (v_n.\bottomright) -- ++(0,-1) node[midway, right] {$e_n$};
	\draw (v_n_3.\bottomright) -- ++(0,-1) node[midway, right] {$e_{n+3}$};
\end{scope}

\end{tikzpicture}
 	\caption{Addition of a new block, consisting of three edges and one vertex, to the existing traffic network. The dashed edges represent not explicitly modeled road links.}
	\label{fig:Network}
\end{figure}

We consider an easily extensible network to show the applicability of our proposed algorithm modifications to more than three dimensions. 
The considered $n$-dimensional traffic network is depicted in \cref{fig:Network} on the left side of the big plus sign.
The dashed arrows represent road links, which exit the network and are not explicitly modeled.
Since we want to evaluate the performance of our approach with respect to the dimensions of the state and input space, we consecutively add a new block to the existing, $n$-dimensional network as shown in \cref{fig:Network}. 
Each newly added block consists of three edges and one vertex, \ie, three road links and one signalized intersection.
Thus, the dimension of the state space in \cref{eq:DimStateSpace} is increased by $3$.
The input space grid doubles, since we assume that every vertex actuates either the incoming arterial road edge or the two intersecting collector edges at \timestep. 

The parameters of the $n$-dimensional network in \cref{fig:Network} are as follows. 
The supply ratio~$\alpha_{e_i, e_j}$ is equal $1$ and the split ratio~$\beta_{e_i, e_j}$ is $0.8$ if $i + 3 = j$ for $j \in \{ 4, 7, \ldots, n-2\}$.
Otherwise, both ratios are equal $0.5$. 
The maximum vehicular capacity $\bar{x}_e$ is $10$ for all edges, and every dimension is quantized in $10$ equally spaced intervals.
Therefore, the considered $n$-dimensional network has $10^n$ state space grid points.
For all arterial road edges, the upper disturbance bound $\bar{d}_e$ is $0$ and the saturation flow $c_e$ is $6$. 
For all intersecting collector edges, the bound is $1$ and the corresponding flow is $2$.

\subsection*{Results}

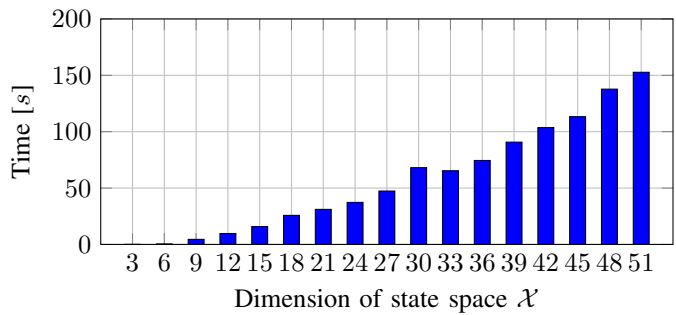
\begin{figure}[htbp]
	\centering
	%\tikzsetnextfilename{AbstractionTime}  
	%
\begin{tikzpicture}
\definecolor{mycolor1}{rgb}{0,0,1}
\setlength\figureheight{3cm} 
\setlength\figurewidth{8cm}

\begin{axis}[%
width=0.951\figurewidth,
height=\figureheight,
at={(0\figurewidth,0\figureheight)},
scale only axis,
bar shift auto,
xmin=0,
xmax=54,
xtick={3,6 ,9,12,15,18,21,24,27,30,33,36,39,42,45,48,51},
xlabel={Dimension of state space $\cxx$},
ymin=0,
ymax=200,
ylabel={Time [$s$]},
axis background/.style={fill=white},
xmajorgrids,
ymajorgrids
]
\addplot[ybar, bar width=1.5, fill=mycolor1, draw=black, area legend] table[row sep=crcr] {%
3	0.01\\
6	0.40\\
9	4.50\\
12	9.70\\
15	15.85\\
18	25.78\\
21	31.10\\
24	37.28\\
27	47.35\\
30	68.09\\
33	65.36\\
36	74.47\\
39	90.71\\
42	103.61\\
45	113.32\\
48	137.74\\
51	152.711\\
};
\end{axis}
\end{tikzpicture}
 	\caption{Computation time of the finite abstraction.}
	\label{fig:AbstractionTime}
\end{figure}

Since the considered $n$-dimensional traffic network in \cref{fig:Network} is a monotone system, we can use \cref{eqn_monotone_over} as reachable set over-approximation function.
The time needed to compute the finite abstraction with respect to the dimension of the state space is illustrated in \cref{fig:AbstractionTime}. 
In contrast to the exponential growth of the brute force algorithm, our proposed \cref{alg_DT_abs} performs roughly linearly with the network length, although the total number of transitions grows exponentially.
For instance, there are $\num{3.3E+62}$ valid transitions in the $51$-dimensional traffic network.

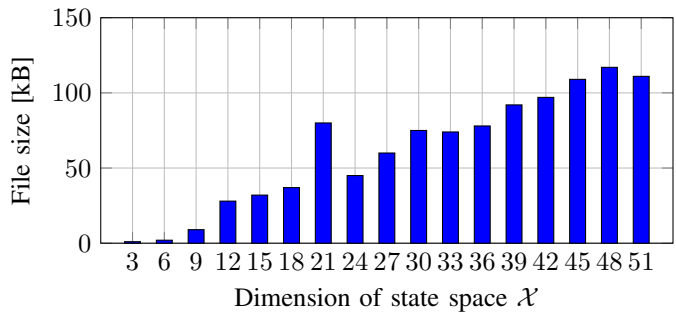
\begin{figure}[htbp]
	\centering
	%\tikzsetnextfilename{AbstractionSize}  
	%
\begin{tikzpicture}
\definecolor{mycolor1}{rgb}{0,0,1}
\setlength\figureheight{3cm} 
\setlength\figurewidth{8cm}

\begin{axis}[%
width=0.951\figurewidth,
height=\figureheight,
at={(0\figurewidth,0\figureheight)},
scale only axis,
bar shift auto,
xmin=0,
xmax=54,
xtick={3,6 ,9,12,15,18,21,24,27,30,33,36,39,42,45,48,51},
xlabel={Dimension of state space $\cxx$},
ymin=0,
ymax=150,
ylabel={File size [\si{\kilo\byte}]},
axis background/.style={fill=white},
xmajorgrids,
ymajorgrids
]
\addplot[ybar, bar width=1.5, fill=mycolor1, draw=black, area legend] table[row sep=crcr] {%
3	1\\
6	2\\
9	9\\
12	28\\
15	32\\
18	37\\
21	80\\
24	45\\
27	60\\
30	75\\
33	74\\
36	78\\
39	92\\
42	97\\
45	109\\
48	117\\
51	111\\
};
\end{axis}
\end{tikzpicture}
 	\caption{File size of the computed transition function BDD.}
	\label{fig:AbstractionSize}
\end{figure}

The file size of the resulting transition function BDD with respect to the dimension of the state space is shown in \cref{fig:AbstractionSize}. 
Due to the memory-efficient data structure, the size of the binary decision diagram which encodes the transition function grows also roughly linearly.

Based on the computed transition function, we can use the fixed point algorithm implemented in \cite{Rungger2016} to synthesize a memoryless, safety supervisory controller for the vehicular traffic network. 
For instance, assume the occupancy of every edge has to be smaller than~$\frac{4}{5}$ of the maximum vehicular capacity for all time steps, \ie in our example less than~$8$.
After representing this safety specification as BDD, we obtain a controller for the $12$-dimensional network after $46$ iterations of the fixed point algorithm.
The synthesis takes $\SI{2.5}{\minute}$ and $\SI{0.9}{\mega\byte}$ are needed to store the obtained controller as binary decision diagram.
This BDD represents all allowed inputs for each discrete state such that the safety specification is fulfilled.
In this example, there exist more than $\num{2.1e+011}$ allowed state-input combinations.

Instead of using the memory-efficient BDD data structure, it is also possible to store the controller as sparse matrix lookup table by explicitly enumerating the state and input space.
However, based on the given dimensions $44$ bits are needed to store a single allowed state-input combination. 
This results in $\SI{1.2}{\tera\byte}$ of memory usage in contrast to $\SI{0.9}{\mega\byte}$ based on the BDD approach.

In contrast to the $12$-dimensional network, it takes $\SI{4.3}{\hour}$ until convergence in the $15$-dimensional case for the same safety specification.
This increase of the controller synthesis computation time has multiple reasons. 
In general, contrary to the coordinate-wise, conjunctive representation of the transition function, a controller cannot be represented in such a compressed form.
Furthermore, the fixed point algorithm determines all allowed state-input combinations.
Since the number of possible control actions for every state grows exponentially with respect to the number of vertices, the controller synthesis problem becomes intractable.
\section{Conclusion}

We have proposed a modification to the abstraction algorithm, which dramatically reduces the computation time for a large class of systems. 
Our traffic example shows that the time and memory requirements for encoding the transition function grows roughly linearly with respect to the network length. 
However, synthesizing controllers for high-dimensional systems still suffers from an exponential runtime with respect to the number of input dimensions. 

Future research will focus on speeding up the synthesis procedure in order to leverage abstraction-based controller synthesis methods.
Additionally, we want to generalize our proposed modifications to continuous-time systems.

\section*{Appendix}

\begin{proof} [\cref{claim_equiv}]

In the following, we show that \cref{alg_brute_abs} and \cref{alg_DT_abs} return the same abstraction.
Without loss of generality, let $n$ be the dimension of the state space~$\cxx$. 
\cref{alg_brute_abs} returns a Boolean function that is of the form
\begin{equation} \label{eqn_BFA_output}
	\bigvee_{ \mathclap{ \substack{\qs \in \qss \\ \qi \in \qis}} } \quad \left( \tovars(\qss) == \tobits(\qs) \wedge \tovars(\qis) == \tobits(\qi) \wedge \left( \bigvee_{\qsp \in \overapprox( \qs, \qi )} \tovars(\qssp) == \tobits(\qsp) \right) \right) .
\end{equation}

According to \cref{ass_over_approx}, the over-approximation function~$\overapprox$ can be decomposed coordinate-wise.
As a result, the next state term $\bigvee_{\qsp \in \overapprox( \qs, \qi )} \tovars(\qssp) == \tobits(\qsp)$ can be split into $n$ components, \ie \cref{eqn_BFA_output} can be written equivalently as
\begin{equation} \label{eqn_split_coord}
	\bigvee_{ \mathclap{ \substack{\qs \in \qss \\ \qi \in \qis}} } \quad \left( \tovars(\qss) == \tobits(\qs) \wedge \tovars(\qis) == \tobits(\qi) \wedge \left( \bigwedge_{i=1}^n \left( \bigvee_{\qsp \in \overapprox( \qs, \qi )} \tovars(\qsspwi) == \tobits(\qspwi) \right) \right) \right) .
\end{equation}
The Boolean expression~\cref{eqn_split_coord} is equivalent to 
\begin{equation} \label{eqn_coordinate_fxn}
	\bigwedge_{i=1}^n \left( \quad \bigvee_{ \mathclap{ \substack{\qs \in \qss \\ \qi \in \qis}} } \quad \left( \tovars(\qss) == \tobits(\qs) \wedge \tovars(\qis) == \tobits(\qi) \wedge \left(  \bigvee_{\qsp \in \overapprox( \qs, \qi )} \tovars(\qsspwi) == \tobits(\qspwi) \right) \right) \right) .
\end{equation}
Equivalence of \cref{eqn_split_coord} and \cref{eqn_coordinate_fxn} can easily be seen when \cref{eqn_coordinate_fxn}'s conjunction over $i$ is viewed as multiplication and the disjunction over $\qs \in \qss$ and $\qi \in \qis$ is viewed as addition. 
``Cross-terms" which correspond to different pairs $( \qs, \qi ) \neq ( \qs', \qi' )$ must equate to Boolean false.

Observe that the output of~\cref{alg_DT_abs} and Boolean function~\cref{eqn_coordinate_fxn} are both conjunctions over coordinates $i \in \{1,\ldots, n\}$. 
If for each coordinate the clauses are equivalent then \cref{alg_brute_abs} and \cref{alg_DT_abs}'s outputs are equivalent, \ie both algorithms return the same abstraction.  

Consider the \ith~component of the system of update equations $\system$ which depends only on a subset of variables. 
The associated box partition projections of the update equation $f_i$ onto the lower dimensional domains are given by $\proj (\qss)$ and $\proj (\qis)$, respectively.
Any state and input variables that do not lie in these two domains do not change the over-approximation of $f_i$. 
Therefore, any of these non-dependent variables with corresponding indices can removed from \cref{eqn_coordinate_fxn} without changing the Boolean expression. 
This removal results in 
\begin{equation} 
	\bigwedge_{i=1}^n \left( \qquad \bigvee_{ \mathclap{ \substack{\qs \in \proj(\qss) \\ \qi \in \proj(\qis)}} } \quad \left( \tovars(\proj(\qss)) == \tobits(\qs) \wedge \tovars(\proj(\qis)) == \tobits(\qi) \wedge \left(  \bigvee_{\qspwi \in \overapproxwi( \qs, \qi )} \tovars(\qsspwi) == \tobits(\qspwi) \right) \right) \right) ,
\end{equation}
which is equivalent to \cref{alg_DT_abs}'s output. 
Therefore, \cref{alg_brute_abs} and \cref{alg_DT_abs} return the same abstraction.

\end{proof} 

\bibliographystyle{unsrt}
\bibliography{Bibliography}

\end{document}